\documentclass{du-journals}
\usepackage{amsmath}
\usepackage{amssymb}
\usepackage{graphicx}
\usepackage{fancyhdr}
\journal{Iranian Journal of Astronomy and Astrophysics}

\title{Local stability criterion for self-gravitating disks in modified gravity}
\author[]{Asiyeh Habibi}
\author[]{M. T. Mirtorabi}
\address[]{Physics Department, Alzahra University, Vanak, 1993891176, Tehran, Iran;}

\author[3]{Mahmood Roshan}
\address[3]{Department of Physics, Ferdowsi University of Mashhad, P.O. Box 1436, Mashhad, Iran;}
\begin{document}
\begin{abstract}
We study local stability of self-gravitating fluid and stellar disk in the context of modified gravity theories which predict a Yukawa-like term in the gravitational potential of a point mass. We investigate the effect of such a Yukawa-like term on the dynamics of self-gravitating disks. More specifically, we investigate the consequences of the presence of this term for the local stability of the self-gravitating disks. In fact, we derive a generalized version of Toomre's local stability criterion for diferentially rotating disks. This criterion is complicated than the original one in the sense that it depends on the physical properties of the disk. In the case of MOdified Gravity theory (MOG), we use the current confirmed values for the free parameters of this theory, to write the generalized Toomre's criterion in a more familiar way comparable with the Toomre's criterion. This generalized Toomre's criterion may be used to study the global stability of stellar and fluid disks using computer simulations.
\end{abstract}

\begin{keywords}
Modified gravity(MOG), Toomre's local stability criterion
\end{keywords}

\section{Introduction}
It is known that self-gravitating stellar disks are strongly unstable to a bar-like mode. In fact, a rotationally dominated disk galaxy can not be a gravitationally stable configuration in the context of Newtonian dynamics. In other words, such a disk rapidly evolves to a pressure dominated system. However, in reality spiral galaxies are stable configurations. This is an evident inconsistency between theory and observation. If we assume that there is no relativistic effect in the evolution of disk galaxies, then we can conclude that for these systems General relativity's predictions are the same as those of Newtonian gravity. Thus, we can say that global instability of disk galaxies is also a problem in General relativity. 

It is well understood in the Astrophysical literature that this inconsistency is linked to the dark matter problem (or the mass discrepancy) in these galaxies. In other words, this inconsistency is similar to that of the observed flat rotation curves of spiral galaxies which are completely different from what expected from the Newtonian gravity. However, by assuming a dark matter halo around the spiral galaxy, both of these inconsistencies can be addressed. It is interesting to mention that the phrase "dark halo" for the first time introduced by Ostriker and Peebles in order to solve the global instability problem of disk galaxies \cite{ostriker}.

Our purpose in this paper is to consider the stability of disk galaxies in the context of some special class of modified gravity theories. One of the main motivations for introducing new modified gravity theories is to solve the dark matter problem. In these theories there are no exotic dark matter particles. So, inconsistencies such as the global instability of disk galaxies should be resolved without any need  to dark matter halos. Now the question is: is it possible to overcome this problems just by properly modifying the gravitational law? 

We consider the stability of disk galaxies in the context of theories which, in the Newtonian limit, add a Yukawa-like term to the Newtonian gravitational potential of a point mass i.e.
\begin{equation}
\phi=-\frac{GM}{r}(c_{1}+c_{2}e^{-\mu r})
\label{petan}
\end{equation}
where $c_1>0$, $c_2$ and $\mu>0$ are arbitrary constants, $G$ is the gravitational constant and $M$ is the mass of the point particle. The numerical values of these parameters $(c_1, c_2, \mu)$ will be determined by the relevant observations. For well-known examples for such theories, we refer the reader to the so called non-local gravity developed by Mashhoon \cite{mashhoon} and Sanders \cite{sanders} and Stelle models \cite{stelle}. Modified Gravity (MOG) is an another example \cite{MOG}. The weak field limit of this theory together with the application of this theory to the spiral galaxies rotation curves and mass discrepancy in galaxy clusters has been studies in \cite{new1,new2}. This theory is one of the current alternative theories to dark matter particles. It is obvious that although the above mentioned theories predict a similar form for the gravitational potential of the point mass, however, in principle, they can lead to different dynamics for self-gravitating systems.

In MOG the parameters $c_1$ and $c_2$ are given by \cite{mog2}
\begin{equation}
c_1=1+\alpha ~~~~~~~~~~~~ c_2=-\alpha 
\end{equation}
where $\alpha$ and $\mu$ are
\begin{equation}
\alpha=\frac{M}{(\sqrt{M}+E)^2}\left(\frac{G_{\infty}}{G}-1\right)~~~~~\mu=\frac{D}{M}
\label{naza1}
\end{equation}
and
\begin{equation}
\begin{split}
& E\simeq  25000 \sqrt{M_{\odot}}\\&
D\simeq  6250 \sqrt{M_{\odot}}kpc^{-1}\\&
G_{\infty}\simeq  20 G
\end{split}
\label{naza2}
\end{equation}
where $M_{\odot}$ is the solar mass. Also in this theory, for an extended system with density $\rho$, in the Newtonian limit, one can verify that the gravitational potential $\Phi=\varphi+\psi$ is determined by the following modified Poisson's equations \cite{mog2} 

\begin{equation}
\nabla^2 \varphi=4\pi G_1 \rho
\label{poas1}
\end{equation}
\begin{equation}
(\nabla^2-\mu^2) \psi=4\pi G_2 \rho
\label{poas2}
\end{equation}
where
$G_1=c_1 G$
and
$G_2=c_2 G$.

The layout of the paper is the following. In section \ref{fluid}, we study the local stability of a fluid disk and find Toomre-like criteria. Section \ref{stellar} is devoted to deriving the local stability criterion for a stellar disk. Discussion is drawn in section \ref{discus}.

\section{Local stability criterion for a fluid disk}
\label{fluid}

In this section, we explore the effect of the Yukawa-like correction term introduced into the gravitational potential on the dynamics of self-gravitating fluid disks. In fact, we find the local stability criterion for a fluid disk. For describing the dynamics of a fluid system, three equations are needed: the continuity equation, the Euler's equation and the modified Poisson's equation (note that in MOG the standard Poisson's equation is not valid). For the sake of simplicity, we assume that the unperturbed disk is axisymmetric and its plane corresponds to $z = 0$ plane. Using the cylindrical coordinates $(R,\phi, z)$, the continuity equation takes the form
\begin{equation}
\frac{\partial\Sigma}{\partial t}+\frac{1}{R}\frac{\partial}{\partial R}\left(\Sigma R v_{R}\right)+\frac{1}{R}\frac{\partial}{\partial\phi}\left(\Sigma v_{\phi}\right)=0
\label{cont}
\end{equation}
where $\Sigma$ is the surface density and $v_{R}$ and $v_{\phi}$ are the velocity components. The components of the Euler's equation can be written as
\begin{equation}
\begin{split}
&\frac{\partial v_{R}}{\partial t}+v_{R} \frac{\partial v_{R}}{\partial R}+\frac{v_{\phi}}{R} \frac{\partial v_{R}}{\partial \phi}-\frac{v_{\phi}^{2}}{R}=- \frac{\partial}{\partial R}\left(\Phi+h\right)\\&
\frac{\partial v_{\phi}}{\partial t}+v_{R} \frac{\partial v_{\phi}}{\partial R}+\frac{v_{\phi}}{R} \frac{\partial v_{\phi}}{\partial \phi}+\frac{v_{\phi} v_{R}}{R}=- \frac{1}{R}\frac{\partial}{\partial\phi}\left(\Phi+h\right)
\end{split}
\label{euler}
\end{equation}
in which $\Phi=\varphi+\psi$ is the gravitational potential and 
\begin{equation}
h= \int \frac{dp}{\Sigma} \ \ \ \ \ \ \ \ c_{s}^{2}=\frac{\partial p}{\partial \Sigma}
\end{equation}
also the corresponding equations for the Poisson's equation are given by \eqref{poas1} and \eqref{poas2}.
Now, we perturb the disk as follows: $\Sigma=\Sigma_{0}+\Sigma_{1}$, $v_{R}=v_{R0}+v_{R1}=v_{R1}$, $v_{\phi}=v_{\phi 0}+v_{\phi 1}$, $\Phi=\Phi_{0}+\Phi_{1}$ and $h=h_{0}+h_{1}$, in which the subscripts "0" and "1" refer to the unperturbed and perturbed quantities respectively. Keeping only terms linear in perturbations in equations (\ref{cont}) and (\ref{euler}), we have 
\begin{equation}
\begin{split}
 &\frac{\partial \Sigma_{1}}{\partial t}+\frac{1}{R}\left(\Sigma_{0} R v_{R1}\right)+\Omega \frac{\partial \Sigma_{1}}{\partial \phi}+\frac{\Sigma_{0}}{R} \frac{\partial v_{\phi 1}}{\partial \phi}=0 \\&
\frac{\partial v_{R1}}{\partial t}+\Omega \frac{\partial v_{R1}}{\partial \phi}-2\Omega v_{\phi 1}=-\frac{\partial}{\partial R}\left(\Phi_{1}+h_{1}\right)\\ &
\frac{\partial v_{\phi 1}}{\partial t}+\Omega \frac{\partial v_{\phi 1}}{\partial \phi}+\frac{\kappa^{2}}{2\Omega} v_{R1}=-\frac{1}{R}\frac{\partial}{\partial \phi}\left(\Phi_{1}+h_{1}\right)
\end{split}
\label{new}
\end{equation}
where $\Omega(R)$ is the circular frequency and the epicyclic frequency $\kappa$ is defined as
\begin{equation}
\kappa^{2}(R)=R \frac{d\Omega^{2}}{dR}+4 \Omega^{2}
\end{equation}
and equations \eqref{poas1} and \eqref{poas2} can be written as

\begin{equation}
\nabla^2\varphi_1 =4\pi G_1 \Sigma_1 \delta(z)
\label{ppoas10}
\end{equation}
\begin{equation}
(\nabla^2-\mu^2)\psi_1 =4\pi G_2 \Sigma_1 \delta(z)
\label{ppoas20}
\end{equation}
This set of equations \eqref{new}-\eqref{ppoas20}, in principle, enable us to analyze the evolution of a given small perturbation. Let us assume the following tightly wound spiral pattern as a perturbation to the background surface density near a point $(R_0,\phi_0)$ (for more detail for such a perturbation see \cite{binney}) 

\begin{equation}
\Sigma_1\simeq\Psi(R_0)~e^{ik(R_0)R}~e^{i(m\phi -\omega t)}
\label{spert}
\end{equation}
It should be noted that we are looking for a local stability criterion at an arbitrary point $(R_0,\phi_0)$ . In the context of Newtonian gravity this criterion is the so-called Toomre's stability criterion \cite{toomre}. Perturbation \eqref{spert} in the surface density causes the perturbation $\Phi_1$ in the gravitational potential as
\begin{equation}
\Phi_1 =\Phi_a\left(R\right)~ e^ {i\left(m\phi -\omega t\right)}
\end{equation}
where
$\Phi_a(R)$
is defined as
\begin{equation}
\Phi_a(R)=\varphi_a(R)+\psi_a(R)
\end{equation}
For tightly wound perturbation, one can neglect the variations with angle $\phi$ since these are much slower than radial variations. Thus we can write
 \begin{equation}
 \begin{split}
& \varphi_1\simeq\phi_a(R)~e^{i(m\phi_0 -\omega t)}\\&
\psi_1\simeq\psi_a(R)~e^{i(m\phi_0 -\omega t)}
\end{split}
\label{ppoas1}
\end{equation}
substituting these equations into \eqref{ppoas10} and \eqref{ppoas20} we get
 \begin{equation}
\nabla^2\varphi_{a} =4\pi G_1 \Sigma_{a}(R) \delta(z)
\label{ppoas2}
\end{equation}
\begin{equation}
(\nabla^2-\mu^2)\psi_{a} =4\pi G_2 \Sigma_{a}(R) \delta(z)
\label{ppoas3}
\end{equation}
where $ \Sigma_a(R)=\Psi(R_0)~e^{ik(R_0)R}$. $\Sigma_1$ closely resembles a plane wave. Now assume that $\textbf{k}(R_0)$ is in the direction of $x$. Therefore, solution of \eqref{ppoas2}
can be gussed as follows
\begin{equation}
\varphi_{a}(R,z)=\varphi_0~ e^{i k R}~e^{-|\alpha' z|}
\label{ppoas4}
\end{equation}
in which
$\alpha'$
and
$\varphi_0$
are arbitrary constants. Assume that there is no matter outside the disk i.e.
$\nabla^{2}\varphi_a=0$. It is easy to show that
$\alpha'=\pm k$
. 
On the other hand at
$z=0$
i.e. where the matter is located, derivative of
$\varphi_a$
with respect to
$z$
is not continuous and consequently the right hand side of 
\eqref{ppoas2}
does not vanish. We integrate from  
\eqref{ppoas2}
$z=-\xi$ to $z=+\xi$
and then let 
$\xi\rightarrow 0$. The result is
\begin{equation}
\varphi_0=\frac{-2\pi G_1}{|k|}\Psi(R_0)
\end{equation}
therefor there is a relation between
$\varphi_a$
and
$\Sigma_a$
as follows
\begin{equation}
\varphi_a(R)=\frac{-2\pi G_1}{|k|}\Sigma_a(R)
\label{ppoas5}
\end{equation}
Now, let us consider equation
\eqref{ppoas3}. Taking into account the simple form of 
$\Sigma_a(R)$
, we can guess the solution for 
$\psi_a(R,z)$
as follows
\begin{equation}
\psi_a(R,z)=\psi_0~ e^{ik R}~e^{-|\alpha' z|}
\label{ppsi}
\end{equation}
for outside the disk, one can verify that
$\alpha'=\pm\sqrt{k^2+\mu^2}$
. Again, derivative of  
\eqref{ppsi} with respect to $z$
at
$z=0$
is not continuous and it is straightforward to show that
\begin{equation}
\psi_0 =\frac{-2\pi ~ G_2}{|\alpha' |}~\Psi(R_0)
\end{equation}
therefor the solution of
\eqref{ppsi}
on the disk is given by
\begin{equation}
\psi_a(R) =\frac{-2\pi ~ G_2}{|\alpha' |}~\Sigma_a(R)
\label{psi}
\end{equation}
Now using
\eqref{ppoas5}
and
\eqref{psi}
, we can write
\begin{equation}
\Phi_a(R)=\frac{-2\pi ~G^*}{|k^*|}\Sigma _a(R)
\label{phisig}
\end{equation}‌
in which
\begin{equation}
 \begin{split}
& k^*=k~\sqrt{k^2+\mu ^2}\\&
G^*=G_1~\sqrt{k^2+\mu ^2}+G_2~|k|
\end{split}
\end{equation}

Therefore, the relation between
$\Phi_a(R)$
and
$\Sigma_a(R)$
in the context of modified theories where the gravitational potential of a point mass is given by 
\eqref{petan} and the modified Poisson's equations are given by \eqref{poas1} and \eqref{poas2}
, is determined by
\eqref{phisig}. If we substitute into equation \eqref{new} the trial solutions of the form
\begin{equation}
\begin{split}
&v_{\phi 1}=v_{\Phi a} e^{i(m\phi_0-\omega t)}\\&
v_{R 1}=v_{R a} e^{i(m\phi_0-\omega t)}
\end{split}
\end{equation} 
then it takes the following form
\begin{equation}
  \begin{split}
& i (m\Omega-\omega)\Sigma_a+\frac{1}{R}\frac{d}{dR}(\Sigma_0 R v_{Ra})+\frac{im\Sigma_0}{R}v_{\phi a}=0\\&
v_{Ra}=-\frac{i}{\overline{\omega}}\left[(m\Omega-\omega)\frac{d}{dR}(\Phi_a+h_a)+\frac{2 m\Omega}{R}(\Phi_a+h_a)\right]\\&
v_{\Phi a}=\frac{1}{\overline{\omega}}\left[\frac{\kappa^2}{2\Omega}\frac{d}{dR}(\Phi_a+h_a)+\frac{m(m\Omega-\omega) }{R}(\Phi_a+h_a)\right]
  \end{split}
  \label{paydari4}
 \end{equation}
it should be noted that 
\begin{equation}
h_a=c_s^2 \frac{\Sigma_a}{\Sigma_0}
\label{referee}
\end{equation}
With the aid of equations \eqref{phisig} and \eqref{referee}, it is quite clear from equation \eqref{paydari4} that we have three equations and three unknowns $v_{\Phi a}$, $v_{R a}$ and $\Sigma_a$. In the following we find the necessary condition for existence of non-trivial solutions for this set of equations. In order to do so, keeping in mind that we are analyzing the local stability, one can ignore the variation of 
$\Sigma_0(R)$
in terms of
$R$
in equations \eqref{new} and assume that
$\Sigma_0(R)$
is constant at
$R_0$. 
Also assuming that 
$\Psi(R)$
is a slowly varying function of
$R$, one can easily derive the following dispersion relation 
\begin{equation}
\left(m\Omega -\omega\right)^2=\kappa ^2+k^2~c_s^2-2\pi G^*\Sigma_0~\frac{k^2}{|k^*|}
\label{pashandegi3}
\end{equation}
For the sake of simplicity we assume that
$m=0$, i.e. we restrict ourselves to axisymmetric disturbances.
Also we define 
$\alpha''$
and
$\beta$
as follows
\begin{equation}
\begin{split}
& \alpha'' =2\pi G\Sigma_0 ~c_1\\&
\beta =\frac{c_2}{c_1}
\end{split}
\end{equation}
with these definitions equation
\eqref{pashandegi3}
can be simplified as
\begin{equation}
\omega ^2=\kappa ^2+~k^2~c_s^2-\alpha'' ~[k+\beta\frac{k^2}{\sqrt{k^2+\mu ^2}}]
\label{pashandegi4}
\end{equation}
This dispersion relation is quite different from the standard one in Newtonian gravity \cite{binney}. As is clear from equation \eqref{pashandegi4}, the right hand side is a real function, hence we can say that all modes have real $\omega^2$. If $\omega^2\geq 0$, then $\omega$ is real and the perturbation oscillate with frequency $\omega$, and the mode is stable. On the other hand if $\omega^2<0$, then $\omega= i\gamma$, where $\gamma$ is a real number, and the mode is said to be unstable. Now, we look for a criterion which ensures the stability of all modes with different wavelengths. To do so, we consider different cases for $\beta$ and find the desired criterion analytically.

\subsection{
Case
$\beta =-1$:
}

The two first terms in the right hand side of equation
\eqref{pashandegi4}
are positive. Therefor the third term's sign is important to determine the sign of 
$\omega^2$. Let us define function
$f(k)$
as follows
\begin{equation}
f(k)=~\left(k-\frac{k^2}{\sqrt{k^2+\mu ^2}}\right)
\label{f(k)}
\end{equation}
\begin{figure}
\centerline{\includegraphics[width=6cm]{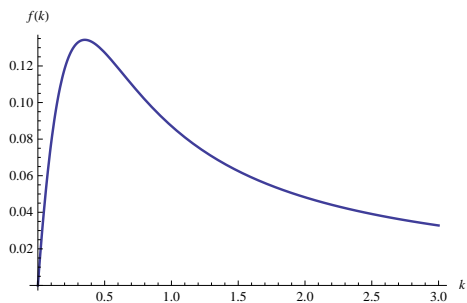}}
\caption[]{
Behavior of
$f(k)$
in terms of $k$
for
$\mu<1$.
}
\label{fk1}
\end{figure}
The behavior of $f(k)$ in terms of $k$ is drown schematically in Figure \ref{fk1}. With this definition, equation
 \eqref{pashandegi4}
can be rewritten as follows
 \begin{equation}
\omega ^2=\kappa ^2+~k^2~c_s^2-\alpha'' ~f(k)
\label{pashandegi5}
\end{equation}
Since
$\alpha''$
is a positive parameter, if we find the wavenumber $k_{max}$ where $f(k)$ is maximum and show that for that wavenumber $\omega^2\geq 0$, then we can be sure that $\omega^2$ is positive for all wavelengths. One can easily show that the maximum value of $f(k)$ is
 \begin{equation}
f(k_{max})=~\frac{(3-~\sqrt{5})~\mu}{\sqrt{2(1+~\sqrt{5})}}
\label{f(k)}
\end{equation}
where
$k_{max}$
is
 \begin{equation}
k_{max}=\sqrt{\frac{\sqrt{5}-1}{2}}~\mu
\label{k}
\end{equation}
Therefore, the system will be stable against all perturbation modes provided that
\begin{equation}
\kappa ^2+c_s^2~k_{max}^2-\alpha'' f(k_{max})> 0
\label{123}
\end{equation}
substituting 
$k_{max}$
and
$f(k_{max})$
into equation \eqref{123} we find
\begin{equation}
\kappa ^2+c_s^2~\left(\frac{\sqrt{5}-1}{2}\right)~\mu ^2-\alpha''\left(\frac{3-\sqrt{5}}{\sqrt{2(1+\sqrt{5})}}\right)~\mu >0
\label{mu}
\end{equation}
this inequality is satisfied for every $\mu<1$ if
\begin{equation}
\frac{c_s\kappa}{\pi G\Sigma_0}>0.472c_1
\label{toomre1}
\end{equation}
This is the local stability criterion for a fluid disk for the case 
$\beta =-1$.

\subsection{
Case
$\beta <-1$:
}
\begin{figure}
\centerline{\includegraphics[width=6cm]{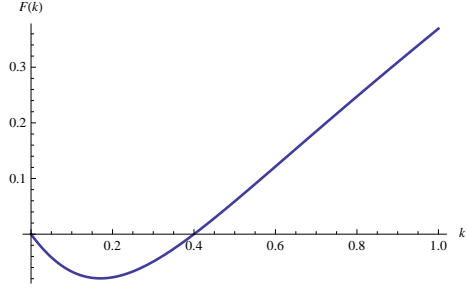}}
\caption[]{
Behavior of
$F(k)$
in terms of
$k$
for
$\mu<1$
and
$\beta<-1$.
}
\label{fk2}
\end{figure}

In this case, let us define
$F(k)$
as follows
\begin{equation}
F(k)=-\left(k+\beta \frac{k^2}{\sqrt{k^2+\mu^2}}\right)
\end{equation}
we have shown schematically the behavior of this function in terms of $k$ in Figure \ref{fk2}.
. We rewrite equation
\eqref{pashandegi4}
as follows
\begin{equation}
\omega^2=\kappa^2 + k^2 c_s^2+\alpha'' F(k)
\label{dovomi}
\end{equation}
Again the sign of 
$F(k)$
will determine the sign of
$\omega^2$
. It is obvious from Figure
\ref{fk2}
that
$F(k)$
has a minimum at
$k=k_{min}$
. It is straightforward to show that
$k_{min}$
is given by
 \begin{equation}
 k_{min}=\left(\frac{3-4\beta^2}{3(\beta^2-1)}-\frac{2\beta\sqrt{4\beta^2-3}}{3(\beta^2-1)}\cos \left(\frac{\theta}{3}\right)\right)^{\frac{1}{2}}\mu
 \end{equation}
 in which
 \begin{equation}
 \theta=\tan^{-1}\left(\frac{3\sqrt{3}\sqrt{-27+86\beta^2-91\beta^4+32\beta^6}}{27-45\beta^2+16\beta^4}\right)
 \label{tet}
 \end{equation}
note that the argument of the radical in equation \eqref{tet} is positive for
 $\beta<-1$
 .
In order to find
 $F(k_{min})$
, for simplicity, we define $f(\beta)$ as 
$k_{min}=f(\beta)\mu$
therefore we can write
 \begin{equation}
 F(k_{min})=-\mu f(\beta)\left(1+\frac{\beta f(\beta)}{\sqrt{1+f(\beta)^2}}\right)
 \end{equation}
if
 $\omega^2$
in
 \eqref{pashandegi4} is positive
for
 $k=k_{min}$, then the system is stable against all perturbation modes with arbitrary wavelengths. Thus a sufficient condition for stability is 
 \begin{equation}
 \kappa^2 +f(\beta)^2+\mu^2 c_s^2-\alpha f(\beta)\left(1+\frac{f(\beta)}{\sqrt{1+f(\beta)}}\right)\mu >0
 \label{fat}
 \end{equation}
it easy to show that the inequality \eqref{fat} is satisfied provided that
\begin{equation}
\frac{c_s \kappa}{\pi G \Sigma_0}>c_1 |1+\frac{f(\beta)}{\sqrt{1+f(\beta)^2}}|
\label{toomre10}
\end{equation}
This is the generalized Toomre's stability criterion. As expected, this criterion is quite different from the standard Toomre's stability criterion in Newtonian gravity.
\begin{figure}
\centerline{\includegraphics[width=6cm]{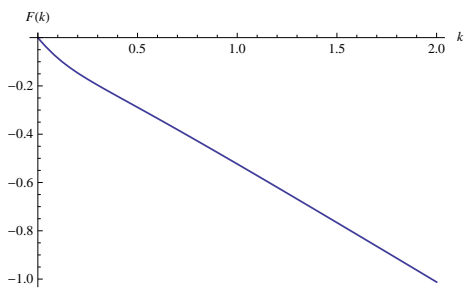}}
\caption[]{
Behavior of 
$F(k)$
in terms of
$k$
for
$\mu<1$
and
$\beta>-1$.
}
\label{fk3}
\end{figure} 
  \subsection{
Case 
$\beta>-1$:  
  }
In this case, the function   
$F(k)$
is drown in Figure
\ref{fk3}.
It is clear from this figure that at large wavenumbers $F(k)$ behaves like a linear function of $k$. Also there is no extremum for this function, and if $k$ increases then  $|F(k)|$ increases. For large wavenumbers we can write
 \begin{equation}
 |F(k)|\simeq (1+\beta)|k|
 \end{equation}
Therefore equation
\eqref{dovomi}
can be rewritten as follows
\begin{equation}
\omega^2\simeq \kappa^2+c_s^2 k^2-2\pi G \Sigma_0 (c_1+c_2)|k|
\end{equation}
now, for stability, the right hand side of the above equation should be positive. Finally, the local stability criterion can be written as
\begin{equation}
\frac{c_s \kappa}{\pi G \Sigma_0}>|c_1+c_2|
\label{tommog}
\end{equation}

\section{Local stability criterion for a stellar disk}
\label{stellar}
\begin{figure}
\centerline{\includegraphics[width=6cm]{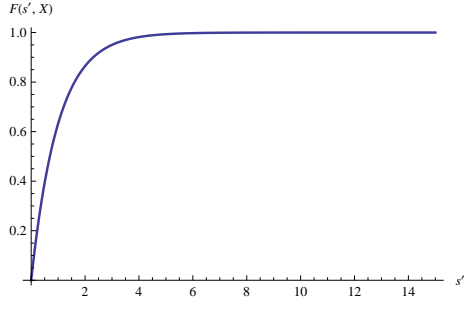}}
\caption[]{$\mathcal{F}(s',\chi)$
in terms of
$s'$
for a constant
$\chi$.
}
\label{redred}
\end{figure}

\begin{figure}
\centerline{\includegraphics[width=6cm]{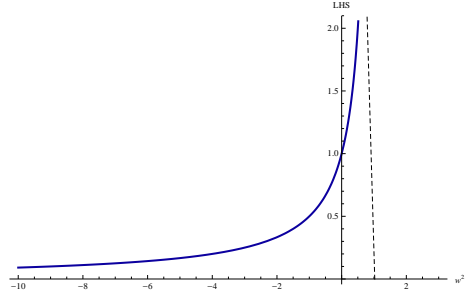}}
\caption[]{
Left hand side of equation
 \eqref{pash3}
in terms of
$\omega^2$.
}
\label{chap}
\end{figure}

Here we study the effect of the Yukawa-like term introduced into the gravitational potential on the local stability of a collisionless stellar self-gravitating disk. The governing equations for such a system are the modified Poisson's equation \eqref{poas1}-\eqref{poas2} and the collisionless Boltzmann equation
  \begin{equation}
 \frac{\partial f}{\partial t}+\textbf{v}\cdot\nabla f-\nabla\Phi\cdot\frac{\partial f}{\partial \textbf{v}}=0
 \label{bolt}
 \end{equation}
 where  $f(\textbf{r},\textbf{v},t)$ is the distribution function. The integration of the distribution function over all velocities gives the matter density $ \rho(\textbf{r})$. Using the Boltzmann equation \label{bolt} in cylindrical coordinate system, one can easily verify the following equations (for more detail see \cite{binney})
\begin{equation}
 \begin{split}
 & \frac{\partial \rho}{\partial t }+\frac{1}{R}\frac{\partial}{\partial R}(R \rho \overline{v_R})+\frac{\partial}{\partial z }(\rho \overline{v_z})=0\\&
 \frac{\partial(\rho \overline{v_R})}{\partial t}+\frac{\partial}{\partial R}(R \overline{v_R}^2)+\frac{\partial}{\partial z }(\rho \overline{v_R}~ \overline{v_z})+\rho\left(\frac{\overline{v_R^2}-\overline{v_{\phi}^2}}{2}+\frac{\partial \Phi}{\partial R}\right)=0\\&
 \frac{\partial(\rho \overline{v_{\phi}})}{\partial t}+\frac{\partial}{\partial R}(\rho \overline{v_R}~\overline{v_{\phi}})+\frac{\partial}{\partial z }(\rho \overline{v_{\phi}}~ \overline{v_z})+\frac{2\rho}{R}\overline{v_{\phi} v_{R}}=0\\&
 \frac{\partial(\rho \overline{v_{z}})}{\partial t}+\frac{\partial}{\partial R}(\rho \overline{v_R}~\overline{v_{z}})+\frac{\partial}{\partial z }(\rho \overline{v_z}^2)+\frac{\rho}{R}\overline{v_{z} v_{R}}+\rho\frac{\partial\Phi}{\partial z}=0
 \end{split}
 \label{bol2}
 \end{equation}
 where
 \begin{equation}
\overline{v_i}=\frac{\int v_i f d^3 v }{\int f d^3 v}=\frac{1}{\rho}\int v_i f d^3 v
\end{equation}
\begin{equation}
\sigma_{ij}^2=\overline{(v_i-\overline{v_i})(v_j-\overline{v_j})}=\overline{v_i v_j}-\overline{v_i}~ \overline{v_j} 
\end{equation}
For a tightly wound spiral perturbation, quite similar to the self-gravitating fluid disk, by linearizing the modified Poisson's and Boltzmann equations and keeping in mind that for a stellar disk $\overline{v_{Ra}}$ is given by (see \cite{binney})
 \begin{equation}
 \overline{v_{Ra}}=\frac{m \Omega-\omega}{\overline{\omega}}k \Phi_a \mathcal{F}
 \label{rf}
 \end{equation}
one can verify the following dispersion relation
 \begin{equation}
(m\Omega-\omega)^2=\kappa^2-2\pi k^2\Sigma_0 \frac{G^*}{|k^*|}\mathcal{F}(\frac{\omega-m \Omega}{\kappa},\frac{k^2 \sigma_R^2}{\kappa^2})
\label{pash1}
\end{equation}
where $\mathcal{F}$ is the reduction factor \cite{binney} and $\sigma_R=\sqrt{\bar{v_R^2}-\bar{v_
R}^2}$ is the radial dispersion velocity. Since the calculations are the same as those of the previous sections, we have neglected them and written down only the results. However it should be noted, that we have not yet specified the reduction factor $\mathcal{F}$. In principle, the form of this function is theory dependent and it could vary from theory to theory. In Appendix 6.A of \cite{binney}, the derivation of the reduction factor in Newtonian gravity has been extensively explored. We repeated this calculations and found out that the reduction factor's form does not change. In other words, $\mathcal{F}$ is the same as of Newtonian gravity and given by
\begin{equation}
\mathcal{F}(s,\chi)=\frac{1-s^2}{\sin\pi s}\int_{0}^{\pi}e^{-\chi(1+\cos\tau)}\sin s\tau ~\sin\tau d\tau
\label{redfac}
\end{equation}
Now we restrict ourselves to the axisymmetric disturbances ($m=0$). In this case, using 
\eqref{pash1}
we get
\begin{equation}
\omega^2=\kappa^2-2\pi G \Sigma_0 c_1\left(|k|+\frac{c_2 k^2}{c_1 \sqrt{k^2+\mu^2}}\right)\mathcal{F}(\frac{\omega}{\kappa},\frac{k^2 \sigma_R^2}{\kappa^2})
\label{pash2}
\end{equation}
For a given wavelength (or mode), if
$\omega^2<0$
then the mode is unstable. Therefore we expect that the system is stable against all axisymmetric perturbation modes if there is no solution of equation \eqref{pash2} with
$\omega^2<0$. In order to check this expectation, we need to study the behavior of
$\mathcal{F}(s,\chi)$
as a function of 
$\omega$. First, let us rewrite 
 \eqref{pash2}
as follows
\begin{equation}
\frac{2\pi G \Sigma_0 c_1}{\kappa^2-\omega^2}\left(|k|+\frac{\beta k^2}{ \sqrt{k^2+\mu^2}}\right)\mathcal{F}(s,\chi)=1
\label{pash3}
\end{equation} 
On the other hand if
$\omega^2<0$
then equation
\eqref{redfac}
can be written as
\begin{equation}
\mathcal{F}(s',\chi)=\frac{1+s'^2}{\sinh (\pi s')}\int _{0}^{\pi}e^{-\chi(1+\cos\tau)}\sinh (s'\tau)\sin\tau d\tau
\label{redfac1}
\end{equation}
in which
$s'=-is$
is a real parameter. For a given wavelength, $\mathcal{F}(s',\chi)$ as a function of
$\chi$
has been shown in Figure
\ref{redred}
.
Note that if
$s'\rightarrow \infty$
(
or equivalently
$\omega^2\rightarrow -\infty$
)
function
$\mathcal{F}(s',\chi)$
for an arbitrary
$\chi$,
asymptotically approaches to $1$. Therefore the left hand side of
\eqref{pash3}
decreases with increasing 
$|\omega^2|$. Also it is maximum at
 $\omega^2=0$ (note that we are considering $\omega^2<0$). The schematic behavior of the left hand side of \eqref{pash3} in terms of
$\omega^2$
is presented in Figure
\ref{chap}. If the left hand side of this equation for
 $\omega^2=0$ is always smaller than $1$, then there is no wavelength with
 $\omega^2<0$ which satisfy the equation
 \eqref{pash3}. In other words, system will be stable to all wavelengths.
Therefore, for
$\omega^2=0$
we expect that
\begin{equation}
\frac{2\pi G \Sigma_0 c_1}{\kappa^2}\left(|k|+\frac{\beta k^2}{ \sqrt{k^2+\mu^2}}\right)\mathcal{F}(0,\chi)<1
\label{pash4}
\end{equation} 
Thus in order to make the stellar disk locally stable, the maximum value of the left hand side of equation \eqref{pash4} should be smaller than $1$. Equation \eqref{pash4} can be rewritten in the form
\begin{equation}
\frac{2\pi G \Sigma_0 c_1}{\kappa\sigma_R}\left(\sqrt{\chi_{max}}+\frac{\beta \chi_{max}}{ \sqrt{\chi_{max}+\left(\frac{\mu\sigma_R}{\kappa}\right)^2}}\right)\mathcal{F}(0,\chi_{max})<1
\label{pashpash}
\end{equation}
where
$\chi_{max}=\frac{\sigma_R^2 k_{max}^2}{\kappa^2}$
and
$k_{max}$
is the wavenumber at which the left hand side of
\eqref{pash4} is maximum. Using the following relation for modified Bessel functions $I_n$
\begin{equation}
e^{z}=I_0(z)+2\sum_{n=1}^{\infty}I_{n}(z)
\end{equation}
and keeping in mind that there is another form for the reduction factor (see \cite{binney}),
  \begin{equation}
 \mathcal{F}(s,\chi)=\frac{2}{\chi} (1-s^2)e^{-\chi}\sum_{n=1}^{\infty}\frac{I_{n}(\chi)}{1-\frac{s^2}{n^2}}
 \label{rrff}
 \end{equation}
 equation
\eqref{pashpash}
reads
 \begin{equation}
 \frac{\kappa\sigma_R}{2\pi G \Sigma_0}>c_1 \left[\left(1+\frac{c_2}{c_1}\frac{\sqrt{\chi}}{\sqrt{\chi+\left(\frac{\mu\sigma_R}{\kappa}\right)^2}}\right)\frac{1-e^{-\chi}I_0(\chi)}{\sqrt{\chi}}\right]_{\chi=\chi_{max}}
 \label{toomtoom}
 \end{equation}
 This is the main result of this section. In fact, equation \eqref{toomtoom} is the generalized version of the standard Toomre's criterion for the local stability of a stellar disk. The standard Toomre's stability criterion is
 \begin{equation}
\frac{\kappa\sigma_R}{3.36 G \Sigma_{0}}>1
\label{toomre2}
\end{equation}
It is important to mention that equation \eqref{toomtoom} is complicated than \eqref{toomre2} in the sense that the right hand is a function of $\sigma_R$ and $\kappa$. In other words, against equation \eqref{toomre2}, the right hand side of \eqref{toomtoom} depends on the physical properties of the disk. We remind that such a dependence is not the case for a fluid disk. Also, it should be stressed that, as expected, the parameter $\mu$ has appeared in the stability criterion.

\begin{figure}
\begin{center}
\includegraphics[width=6cm]{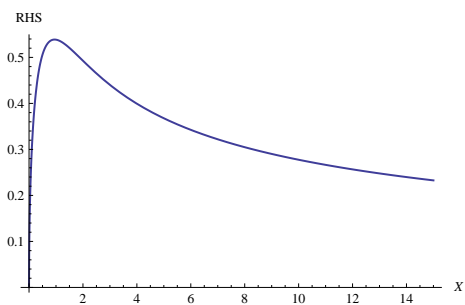}
\caption{
Right hand side of equation
 \eqref{toomtoom}.
}
\label{rast}
\end{center}
\end{figure}

In the case of MOG
 ،
 $c_1=1+\alpha$
and
 $c_2=-\alpha$. It should be noted that it is not analytically possible to find the maximum value of the right hand side of equation \eqref{toomtoom}. Even in Newtonian gravity where
 $c_1=1$
and
 $\beta=0$, one should use numerical calculations in order to find out the maximum value. In this case the maximum value is $0.53449$. Although we are sure that there is a maximum, see Figure \ref{rast}, however we need the numerical values of 
 $\alpha$ \textbf{and $\mu$} to find the maximum value of the right hand side of equation \eqref{toomtoom}.
The current observational values of MOG's free parameters are 
$\alpha=8.89\pm0.34$ and $\mu = 0.042 \pm 0.004~ kpc^{-1}$ \cite{new2}. As we mentioned before, the generalized Toomre's criterion \eqref{toomtoom} depends on $\sigma_R(R)$ and epicycle frequency $\kappa(R)$. For example in the solar neighborhood $\sigma_R=(38\pm2)~km~s^{-1}$ and $\kappa=(37\pm 3)~km~s^{-1} kpc^{-1}$. Substituting these values to the RHS of \eqref{toomtoom}, one can verify that the maximum value of the RHS is $0.728$. In this case the generalized Toomre's criterion can be rewritten as 
\begin{equation}
\frac{\kappa\sigma_R}{4.58 G \Sigma_0}>1
\label{newtoom}
\end{equation}
It is worth to recall that, in the case of Newtonian gravity the numerical coefficient in the denominator is 
$3.36$. This shows that in MOG, for the local stability of a stellar disk, the ratio
$\frac{\kappa\sigma_R}{\Sigma_0}$ should be larger than that of Newtonian gravity.
In other words, if we assume the same values for  
$\Sigma_0$
and
$\kappa$ in both MOG and Newtonian gravity, then a larger dispersion velocity is needed in MOG to establish the local stability.

\section{Discussion}
\label{discus}
In this paper we have studied the local stability of self-gravitating stellar and fluid disks in the context of special class of modified gravity theories. In fact, we considered the local stability in theories which introduce a Yukawa-like term into the gravitational potential of a point mass. We explored the effect of such a term on the local stability of disk galaxies. We showed that, in principle, the local stability criterion for both stellar and fluid disks are different from the corresponding criterion in Newtonian gravity, i.e. Toomre's stability criterion. Our main results for a fluid disk are equations \eqref{toomre1}, \eqref{toomre10}, \eqref{tommog}. These criteria correspond to the cases $\frac{c_2}{c_1}=-1$, $\frac{c_2}{c_1}<-1$ and $\frac{c_2}{c_1}>-1$ respectively. Where $c_1$ and $c_2$ are free parameters of the given theory and appear in the modified Poisson's equations. Furthermore for a stellar disk the stability criterion is given by equation \eqref{toomtoom}. 

As we already mentioned in the introduction, the global instability of the disk galaxies is directly related to the dark matter problem. Our main purpose in this paper was to find the local stability criterion which is necessary for performing computer simulations for studying dynamics and evolution of the disk galaxies \cite{ostriker}. In other words, in order to check that the above mentioned theories can solve the global instability problem of the disk galaxies, we need to study the global stability issue of this systems via computer simulations \cite{us}. However, it is worth to mention that, in MOG with special choices for the free parameters, the generalized Toomre's criterion \eqref{newtoom} is not significantly different from the standard Toomre's criterion. Albeit, one can not conclude that the bar instability exists in MOG. In other words without performing computer simulations we can not decide about the global instability of the self-gravitating disks in MOG just by comparing the local stability criterion with the Toomre's criterion. Studying the global stability of stellar disks in the context of MOG using computer simulations is left as a subject of future study \cite{us}.

\section*{Acknowledgment}
We would like to thank Fatimah Shojai for vluable comments and useful discussions. Also we would like to thank the referee for useful comments that helped us to improve the paper.

\end{document}